\documentstyle[epsf,12pt, bbm]{article}
\newcommand{\be}{\begin{equation}}
\newcommand{\ee}{\end{equation}}
\newcommand{\balpha}{\mbox{\boldmath$\alpha$}}

\newcommand{\bb}{\mbox{\boldmath$b$}}
\expandafter\ifx\csname mathbbm\endcsname\relax

\else

\fi
\begin{document}
\begin{titlepage}
\begin{flushleft}
       \hfill                      USITP-95-12, UUITP-20/95\\
       \hfill                       November 1995\\
\end{flushleft}
\vspace*{3mm}
\begin{center}
{\LARGE Exceptional Equivalences \\
in N=2 Supersymmetric Yang-Mills Theory \\}
\vspace*{12mm}
{\large Ulf H. Danielsson\footnote{E-mail: ulf@rhea.teorfys.uu.se} \\
{\em Institutionen f\"{o}r teoretisk fysik \\

Box 803\\
S-751 08  Uppsala \\
Sweden \/}\\
\vspace*{5mm}
Bo Sundborg\footnote{E-mail: bo@vana.physto.se} \\
        {\em Institute of Theoretical Physics \\
         Fysikum \\
        Box 6730\\
        S-113 85 Stockholm\\
        Sweden\/}\\}
\vspace*{15mm}
\end{center}

\begin{abstract}
We find low energy equivalences between $N=2$ supersymmetric gauge theories
with different simple gauge groups with and without matter.  We give a
construction of equivalences based on subgroups and find all examples with
maximal simple subgroups. This is used to solve some theories with exceptional
gauge groups $G_2$ and $F_4$. We are also able to solve an $E_6$ theory on a
codimension one submanifold of its moduli space.
\end{abstract}

\end{titlepage}

\section{Introduction}

In this paper we will discuss some techniques to solve $N=2$ supersymmetric
Yang-Mills theories. After the solution of $SU(2)$ by Seiberg and Witten
\cite{SW} all theories with non-exceptional gauge groups have been solved with
various matter content [2-12]. We will discuss equivalences between a large
number of  theories and use these to solve some cases with exceptional gauge
groups.  This will be accomplished using only hyperelliptic surfaces.
Let us begin with an overview of some important  concepts.

In $N=1$ super-space the Lagrangian is
given by
\be
{ 1 \over 4 \pi} Im \left( \int d^{4} \theta { \partial {\cal F} (\Phi )
\over \partial \Phi } \bar{\Phi} + \int d^{2} \theta {1 \over 2}
{ \partial ^{2} {\cal F} ( \Phi ) \over \partial \Phi ^{2}} W^{2}
\right)   ,
\ee
where $W = (A,\lambda )$ is a gauge field multiplet and $\Phi =(\phi , \psi )$
is a chiral
multiplet, both taking values in the adjoint representation.
The field $\phi$ is given a vacuum expectation-value according to
\be
\phi = \sum _{i=1}^{r} b_{i} H_{i}   ,\label{Cartan}
\ee
where $r$ is the rank of the group. $H_{i}$ are elements of the
Cartan sub-algebra.
At a generic $\bb$ the gauge group is broken down to
$U(1)^{r}$ and each $W$-boson, one for each root $\balpha$, acquires a
$(mass)^2$
proportional to $(\bb \cdot \balpha )^{2}$.
Restoration of symmetry
(classically) is obtained when $\bb$ is orthogonal to a root.
At such a point the $W$-boson corresponding to that root becomes
massless.

The action above is an action for the massless $U(1)$ gauge fields after that
the massive fields have been integrated out. As discussed in \cite{var}, the
perturbative prepotential in terms
of the $N=2$ superfield $\Psi$ is given by
\be
{\cal F} \sim {i \over 4 \pi}
\sum _{\balpha} (\Psi \cdot \balpha )^2 \log{ (\Psi \cdot \balpha )^2  \over
\Lambda ^2}    ,
\label{F}
\ee
i.e. by one loop contributions only.
Let us now introduce matter in the form of $N=2$ hypermultiplets. We will only
consider hypermultiplets with the bare mass put to zero. These hypermultiplets,
like the vector multiplets, will receive masses through the Higgs mechanism.
The $(mass)^2$  of the hypermultiplets will be proportional to  $ ( {\bf w}
\cdot  \bb )^2$
where $\bf{w}$ is a weight in the representation corresponding to the
hypermultiplet. The full perturbative prepotential is now given by
\be
{\cal F} \sim {i \over 4 \pi}
\sum _{\balpha} (\Psi \cdot \balpha )^2 \log{ (\Psi \cdot \balpha )^2  \over
\Lambda ^2} -
{i \over 4\pi}
\sum _{{\bf w}} (\Psi \cdot {\bf w} )^2 \log{ (\Psi \cdot {\bf w} )^2  \over
\Lambda ^2}    .
\label{verkan}
\ee
The sum over weights $\bf w$ is over all weights of all matter hypermultiplets
with multiplicity.
The full nonperturbative expression is obtained by making sure that the
effective coupling $\tau _{ij} = Im ( \frac{\partial ^{2} {\cal F}
(a)}{\partial  a_i \partial a_j})$ is positive definite. The trick is to
construct a suitable Riemann surface whose period matrix is identified with the
effective coupling. The monodromies obtained from ${\cal F}$ by acting with the
Weyl group should then also be reproduced by the cycles on the Riemann surface.
Let us list some of the results obtained so far using this method.  The
$SU(N_{c})$ curve with $N_{f}$ massless, fundamental hypermultiplets is given
by, \cite{KLYT,APS,HO},
\be
y^2 = ((x-b_{1})...(x-b_{N}))^2 - x^{N_{f}} \Lambda ^{2N_{c} - N_{f}}
\ee
where $\sum _{i=1}^{N_{f}} b_{i} =0$.
For $SO(2r+1)$ with $N_{f}$ massless, fundamental hypermultiplets we have,
\cite{var,H},
\be
y^{2} =((x^2 -b_{1}^{2})...(x^{2}-b_{r}^{2}))^{2} - x^{2 +2N_{f}} \Lambda
^{4r-2-2N_{f}}
\ee
and for  $SO(2r)$ with $N_{f}$ massless, fundamental hypermultiplets we have,
\cite{BL,H},
\be
y^{2} =((x^2 -b_{1}^{2})...(x^{2}-b_{r}^{2}))^{2} - x^{4 +2N_{f}} \Lambda
^{4r-4-2N_{f}}.
\ee
The exponent of $\Lambda$ is given by $I_{2}(R_{A}) - \sum I_{2}(R_{M})$, where
$I_{2}(R_{A})$ is the Dynkin index of the adjoint representation of the vector
multiplet (i.e. twice the dual Coxeter number), while $I_{2}(R_{M})$ is the
Dynkin index of the representation of the matter hypermultiplets.
In all these cases it has been possible to use a hyperelliptic surface.

More general methods of generating solutions were suggested in \cite{DW,MW}. So
far, however, no explicit results have been obtained for the exceptional
groups, although the construction of \cite{MW} hints that in general non
hyperelliptic surfaces might be needed.

In the next section we will discuss the general principles behind equivalences
of  $N=2$ supersymmetric Yang-Mills theories. We will also list a natural class
of relations between theories with simple gauge groups. Subsequent sections
will deal with explicit and illustrative examples.

\section{Equivalences through subgroups}

\subsection{Construction requirements}

We can now exploit the general equation (\ref{verkan}) for the
prepotential to explain some relations between $ N=2 $
gauge theories with different
gauge groups and different matter content. Isolated examples of
the type we  are considering have been observed before in the
literature \cite{AS}, but we now use group theory to survey
systematically where one can take advantage of such relations.

As will be seen explicitly in examples in the following
sections, cancellations between vector multiplet terms and
hypermultiplet terms in the prepotential can often be arranged
so as to give identical prepotentials for different theories.
Then we expect, as all experience to date indicates, that
the semi-classical prepotential determines the full low-energy
solution of the theory by providing the (singular)
boundary conditions at infinity. The analyticity property of the
prepotential seem to make the solutions unique. In any case, if
two additional constraints on the solutions,
Weyl symmetry and anomalous $ U(1)_R $ symmetry, are satisfied we
cannot ask for more. The semi-classical prepotential is constructed
to pass these tests, Weyl symmetry since the weight systems of unitary
representations are Weyl symmetric, and $ U(1)_R $ symmetry because
of an argument given in \cite{DS}.

The essential feature of the prepotential (\ref{verkan}) is that the
terms containing the weights of the adjoint representation (the roots)
appear with positive sign, and terms containing weights of the
matter representations appear with a negative sign. The reason is
that vector multiplets and matter hypermultiplets contribute with
opposite signs to the beta function. Precisely this difference in
signs is at the core of the equivalences we shall study. Namely,
the weight systems of some matter representations may overlap with
and partly over-shadow the root system. While there is a one to one
correspondance between semi-simple groups and root systems, the
effect of a change
of group and root system on the prepotential
may sometimes be compensated by an appropriate matter content.

The simplest instances when weight lattices of different groups
overlap occur when one group $\tilde{G}$ is a
subgroup of the other, $G$. This is
what we shall investigate. We shall find two qualitatively different
cases. If $\tilde{G}$ and $G$ are of equal rank the moduli
spaces for the two theories have the same dimensions. Then, if there
is also a
one to one mapping between the moduli spaces in the
semi-classical region relating the prepotentials
$\tilde{{\cal F}}$ and ${\cal F}$ it will imply a one to one mapping
between the two moduli spaces. At the effective
level the theories are then
equivalent. If on the other hand the rank of $\tilde{G}$ is less
than the rank of $G$, there can at most be an embedding of the
semi-classical moduli space of $\tilde{G}$ into the one of $G$,
relating the prepotentials. In this case the $\tilde{G}$ theory will be
contained in the $G$ theory as a subset of its moduli space.

Group theory tells us how to describe a gauge theory with a given
gauge group in terms of one of its subgroups. For each representation
there is a branching rule which describes it in terms of a sum
of representations of the subgroup. Normally we like to have
vectors in the adjoint representation of the gauge group, but
the adjoint always branches to the adjoint of the subgroup
{\em plus\/} other representations,
so we need to take care of such extra vectors.
In the low-energy
theory with symmetry broken to a product of $U(1)$ factors, it
is enough to consider the contribution of these states to
the effective prepotential.
But terms from the
vector multiplets not in the adjoint can sometimes be
cancelled by hypermultiplet
contributions. We ask in general\footnote{We shall
 also give an example where the subgroup theory
is obtained by a more complicated procedure.}
\begin{eqnarray}
	R_A & \to & \tilde{R}_A + \tilde{R}_0
	\label{AdjointBranching} \\
	R_M & \to & \tilde{R}_0 + \tilde{R}_M
	\quad ,
	\label{MatterBranching}
\end{eqnarray}
for the branching adjoint vectors and matter representations,
respectively.
This branching scheme
ensures a low-energy embedding of the $\tilde{G}$ theory
with an $\tilde{R}_M$ hypermultiplet into the $G$ theory with
an $R_M$ hypermultiplet. Note that we have not required the
matter representations to be irreducible. In general, we
can allow for an arbitrary number of singlet hypermultiplets
in both theories without changing the behaviour of the
prepotentials. The reason is that
neutral couplings between vectors and hypermultiplets
are not allowed by $N=2$ supersymmetry \cite{WLP}. Therefore,
we disregard any singlet hypermultiplets in the following
discussion.

An important invariant of a representation $R$ is its
second order Dynkin index $I_2(R)$.
Since it enters
the one-instanton term $\Lambda^{I_2(R_A) -  I_2(R_M)}$ in
the prepotential, we need to understand
the Dynkin indices of representations of subgroups in order
to see if we get the correct one-instanton contributions
(and anomalous $U(1)_R$ symmetry). The rule is that
\begin{equation}
	I_2(R) = I_{\tilde{G}\subset G} I_2(\tilde{R}) \quad,
	\label{SubgroupIndexDef}
\end{equation}
where $I_{\tilde{G}\subset G}$ is an integer depending only
on the embedding of the subgroup $\tilde{G}$ in $G$. It
follows that we can get unchanged instanton expansions if
\begin{equation}
	I_{\tilde{G}\subset G} = 1 \quad .
	\label{SubgroupIndexEq}
\end{equation}
If this condition is violated it appears that the
one-instanton term is missing from the $\tilde{G}$
reduction of the $G$ theory.

\subsection{Search for subgroups}

We have found that we can expect a simple relation between
$N=2$ gauge theories with groups $G$ and $\tilde{G}\subset G $ if the
index of the subgroup is unity (\ref{SubgroupIndexEq}) and the
branching rules of the matter representations can compensate for
those of the vectors in the adjoint
(\ref{AdjointBranching},\ref{MatterBranching}).
We now proceed to the survey of what Lie subgroups can satisfy these
two conditions.


We restrict our attention to subgroups that are simple, and
to the simple subgroups $\tilde{G}$ that are also maximal, i.e.
such that there are no other simple subgroups $G'$ with
$\tilde{G} \subset G' \subset G $. Given all maximal simple
groups one can build a hierarchy of simple subgroups, and
this can of course also be done for the equivalences we shall
list for $N=2$ theories. However, we
only claim that the list itself is
exhaustive. There may exist non-maximal equivalences which
cannot be obtained by a chain of the maximal equvalences in our list.

The root system of the subgroup can be a subset
of the root system of the full group. Then the subgroup is called
a {\em regular\/} subgroup. Otherwise, we have a {\em special\/}
subgroup. Special subgroups are always of lower rank, but regular
subgroups may have the same rank as the full group. Discussions
of subgroups and useful tables can be found in refs. \cite{D,S,F}.

\begin{figure}
\hbox to\hsize{\hss
\epsfysize=8cm
\epsffile{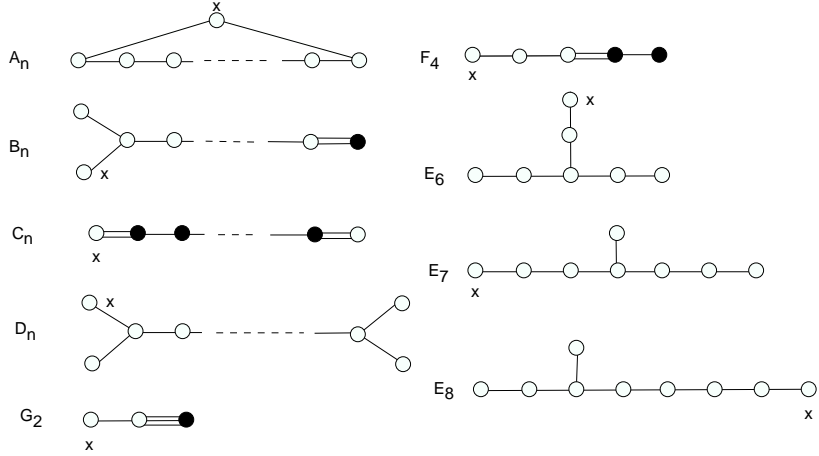}\hss}
\caption{Extended Dynkin diagrams for simple groups. The ordinary Dynkin
diagrams are obtained by removing the roots labelled $x$.}
\end{figure}

Semi-simple maximal regular subgroups
can be read off from the Dynkin diagrams obtained
by deleting a vertex from the so-called extended Dynkin diagram
of a group (fig 1). The final number of vertices is the same as in the
original Dynkin diagram, which means that the rank is preserved.
The only cases which lead to {\em simple\/} subgroups are
\begin{equation}
	\begin{array}{ccccc}
		 & SO(2n) & \subset & SO(2n+1) & R_M = {\bf (2n+1)}
		\label{RegularSeries}  \\
		 & SU(3) & \subset & G_2 & R_M = {\bf 7}
		\label{RegularG_2}  \\
		\tilde{R}_M = {\bf 9} & SO(9) & \subset & F_4 & R_M = {\bf 26}
		\label{RegularF_4} \\
		\ast & SU(8) & \subset & E_7 & \ast
		\label{RegularE_7} \\
		\ast & SO(16) & \subset & E_8 & \ast
		\label{RegularOE_8} \\
		\ast & SU(9) & \subset & E_8 & \ast
		\label{RegularUE_8}
	\end{array} \quad .
\end{equation}
All these examples have Dynkin index $I_{\tilde{G}\subset
G} = 1$. We have written down the matter representations in terms of
their dimensions, but the examples marked by asterisks
unfortunately do not possess suitable
matter representations.

One can also delete one of the vertices from the
original Dynkin diagram. If it is only connected to one other vertex,
the result will be a connected Dynkin diagram, corresponding to a
simple subgroup, which is a maximal simple subgroup if the diagram
is not a sub-diagram of a simple subgroup obtained from the
extended Dynkin diagram. In this way one gets regular
subgroups with rank
reduced by one. The corresponding sub-theories are obtained by
considering only points in the moduli space satisfying
${\bf b \cdot \Lambda}_k = 0$, where ${\bf \Lambda}_k$ is the highest
weight of the basic representation corresponding to the deleted
vertex. (${\bf \Lambda}_k$ is orthogonal to all
remaining roots, so in
effect we get an orthogonal projection on the space spanned by
these roots.) Note the difference of this {\em exact\/} way of
finding subgroups, which only works in special cases, and the
general but approximate subgroups one finds in the semi-classical
regime by strong Higgs breaking \cite{AF,var}. The exact embeddings of theories
we find are listed below, together with candidate subgroups which
lack appropriate matter representations (marked with asterisks).
\begin{equation}
\begin{array}{ccccc}
		& SU(n-1) & \subset & SU(n) & R_M ={\bf n}+{\bf \bar{n}} \\
    	& SO(2n-1) & \subset & SO(2n+1) & R_M = 2\cdot {\bf (2n+1)} \\
	    & Sp(2n-2) & \subset & Sp(2n) & R_M =2\cdot {\bf (2n)} \\
	    & SO(2n-2) & \subset & SO(2n) & R_M =2\cdot {\bf (2n)} \\
\tilde{R}_M =2\cdot {\bf 10} & SO(10) & \subset & E_6 &
R_M={\bf 27}+{\bf \bar{27}} \\
	    & E_6 & \subset & E_7 & R_M ={\bf 56}\\
	    & SU(2) & \subset & G_2 &  R_M =2\cdot {\bf 7}\\
   \ast & SU(n) & \subset & SO(2n) & \ast  \\
	    & SU(4) & \subset & SO(8) &  R_M =2\cdot {\bf 8}_s \\
\tilde{R}_M={\bf 5}+{\bf \bar{5}}
        & SU(5) & \subset & SO(10) &  R_M={\bf 16}+{\bf \bar{16}} \\
	    & SU(6) & \subset & SO(12) &  R_M={\bf 32}' \\
   \ast & SU(n) & \subset & Sp(2n) & \ast  \\
	    & SU(2) & \subset & Sp(4) & R_M=2\cdot {\bf 5}  \\
	    & SU(3) & \subset & Sp(6) & R_M={\bf 14}'  \\
   \ast & SU(6) & \subset & E_6 & \ast \\
   \ast & SO(12) & \subset & E_7 &  \ast \\
   \ast & E_7 & \subset & E_8 & \ast  \\
   \ast & Sp(6) & \subset & F_4 & \ast
\end{array}
\end{equation}
These examples also have $I_{\tilde{G} \subset  G} = 1$. The series
of unitary subgroups of orthogonal and symplectic groups, do not
in general have matter representations giving asymptotically free
theories. Only the special cases that are listed without asterisks
satisfy this requirement. If there are several representations of the same
dimension, we have distinguished them by the notation of \cite{S}.

The special subgroups have been classified by Dynkin \cite{D}. Of
the simple subgroups we again list those which are maximal and have
unit Dynkin index (an effective criterion to rule out possibilities
in this case). Again, embeddings which work are marked by their
matter content, and those which do not have appropriate representations
are marked by asterisks.
\begin{equation}
\begin{array}{ccccc}
	  & SO(2n-1) & \subset & SO(2n) & R_M={\bf 2n}\\
      & Sp(2n) & \subset & SU(2n) & R_M={\bf (n+1)(2n+1)}\\
      & G_2 & \subset & SO(7)  & R_M={\bf 7} {\rm \ or\ } {\bf 8}\\
      & F_4 & \subset & E_6 & {\bf 27} \\
\ast  & G_2 & \subset & F_4 & \ast \\
\ast  & G_2 & \subset & E_6 & \ast\\
\ast  & G_2 & \subset & E_7 & \ast \\
\ast  & G_2 & \subset & E_8 & \ast \\
\ast  & Sp(8) & \subset & E_6 & \ast \\
\ast  & Sp(6) & \subset & E_7 & \ast \\
\ast  & F_4 & \subset & E_7 & \ast \\
\ast  & F_4 & \subset & E_8 & \ast
\end{array}
\end{equation}
To these special group embeddings correspond embeddings of
moduli spaces, just as in the case of the regular embeddings.
However, we cannot give a simple and general description of the
special embeddings.

\section{Some regular examples}

\subsection{$F_4$ with fundamental matter}

The first of the examples that we will study involves the exceptional group
$F_4$.
The adjoint representation of $F_4$ has dimension 52. It has 4 Cartan elements
and 48 roots. Let us add one hypermultiplet in the fundamental $\bf{26}$. This
will effectively cancel all the long roots leaving only the short ones. The
remaining weights  are the roots of  $SO(8)$. What we actually have used is the
regular embedding of $SO(9)$ in $F_4$ and the regular embedding of $SO(8)$ in
$SO(9)$. Hence we can argue that the low energy theory of $F_4$ with one
fundamental matter hypermultiplet is the same as the pure $SO(8)$ theory. The
check of  the Dynkin indices works out as $18-6=12$.

Let us repeat the logic of the argument. Let us assume the existence of the
$F_4$ solution. The discussion above then shows that this solution obeys all
the requirements also of the $SO(8)$ theory (these are less restrictive since
the Weyl group is smaller). Hence it follows, if the $SO(8)$ solution is
unique, that the $F_4$ solution must be identical to the $SO(8)$ solution that
we already know. In fact, the existence of the $F_4$ theory implies certain
symmetries of the $SO(8)$ theory that from the $SO(8)$ point of view look
accidental.

There is something surprising about this equivalence. So far all constructions
of curves for gauge groups with various types of matter have been invariant
under the Weyl group. Indeed, this can be used as an important guide when
finding the curves. More precisely, the weight diagram for the fundamental
representation of the group has been used to construct the curve. This is the
simplest way of finding a representation of the Weyl group. The Weyl group of
$F_4$ does not leave the suggested curve invariant. However, there is still a
way out. The only thing we really need is that the Weyl group is represented on
the integrals over the cycles. This is trivially true at the perturbative
level, but a very strong requirement non-perturbatively.
The reason that this works is the triality symmetry of $SO(8)$. The $SO(8)$
curve is based on the fundamental $\bf{8}$ of $SO(8)$. There are in fact three
equivalent representations and therefore three equivalent curves. The $F_4$
Weyl elements not contained in the $SO(8)$ Weyl group permute these three
different curves.
If we set the projections of the weights $(1 0 0 0)$, $(-1 1 0 0)$, $(0 -1 1
1)$ and $(0 0 -1 1)$ equal to $b_1$,  $b_2$, $b_3$ and $b_4$ respectively,
there is a sequence of $F_4$ Weyl transformations that  takes
$$
b_1 \rightarrow b_1 - \Delta _1  \rightarrow b_1 - \Delta _1 - \Delta _2
\rightarrow b_1
$$
$$
b_2 \rightarrow b_2 + \Delta _1  \rightarrow b_2 +\Delta _1 + \Delta _2
\rightarrow b_2
$$
$$
b_3 \rightarrow b_3 + \Delta _1  \rightarrow b_3 + \Delta _1 + \Delta _2
\rightarrow b_3
$$
\be
b_4 \rightarrow b_4 + \Delta _1  \rightarrow b_4 + \Delta _1 - \Delta _2
\rightarrow -b_4
\ee
where
\be
\Delta _{1} = {1 \over 2} (b_1 - b_2 - b_3 - b_4 ) \;\; \mbox{and} \;\;
\Delta _2 = b_4  .
\ee
This is illustrated in fig. 2. For consistency we must have
 $a_1 ' ={1 \over 2}  (a_1 + a_2 + a_3 + a_4)$ including the non-perturbative
corrections, e.g.,
\be
\oint _{\gamma _{1}'} \lambda ={1 \over 2}  (\oint _{\gamma _{1}} \lambda  +
\oint _{\gamma _{2}} \lambda +\oint _{\gamma _{3}} \lambda + \oint _{\gamma
_{4}} \lambda ),
\label{under}
\ee
where the respective cycles are indicated in fig. 2. We have that $\lambda =
(2p-xp'){dx \over y}$ with $y^2 =p^2 - x^4 \Lambda ^{12}$ and $p= (x^2-b_1
^2)...(x^2 - b_4 ^2)$. These integrals are easily calculated in an expansion in
$\Lambda ^{12}$. It can be checked that (\ref{under}) is really satisfied.

\begin{figure}
\hbox to\hsize{\hss
\epsfysize=10cm
\epsffile{
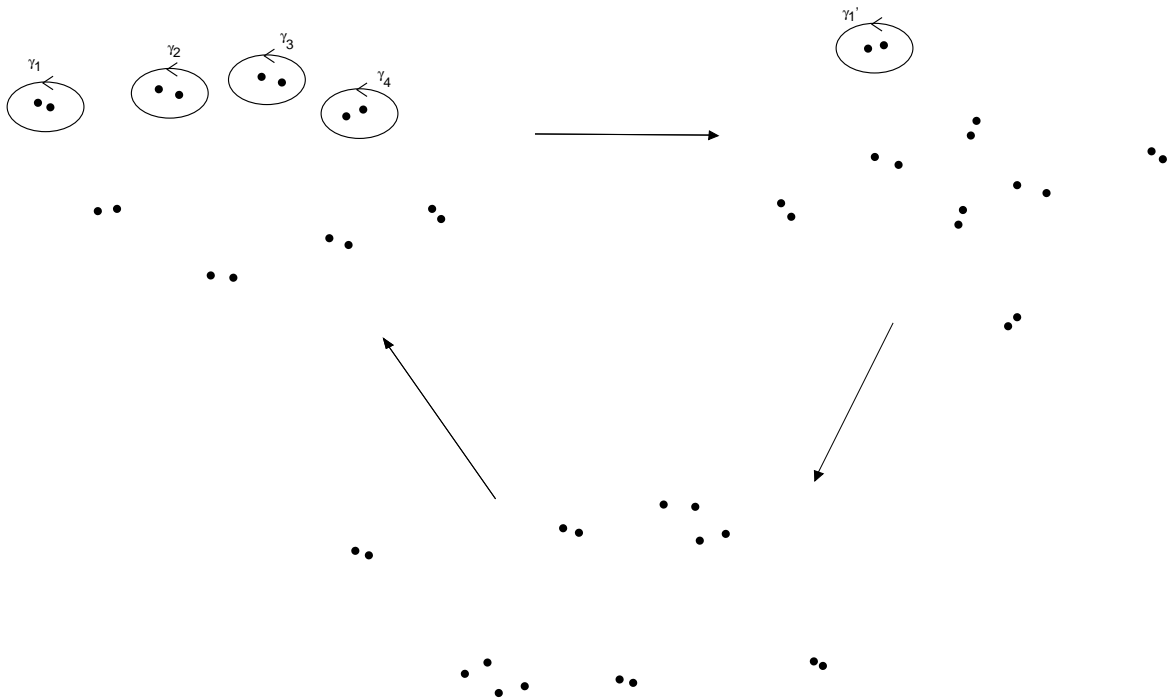}\hss}
\caption{The three curves for $F_4$ with fundamental matter interchanged by
$SO(8)$ triality.}
\end{figure}

{}From this we can conclude that $F_4$ with one massless fundamental
hypermultiplet is given by
\be
y^2 = ((x^2-b_1 ^2)(x^2 - b_2 ^2 )(x^2 - b_3 ^2 )(x^2 - b_4 ^2))^2 - x^4
\Lambda ^{12}
\ee
We emphasize again that the curve can not be written in terms of the $F_4$
Casimirs.

\subsection{ $G_2$ with fundamental matter}

There is an even simpler example of the phenomena discussed above.  Let us
consider $G_2$ with matter in the fundamental $\bf{7}$. The effective low
energy theory can be shown to be the same as that of pure $SU(3)$. The breaking
to $SU(3)$ gives ${\bf 14} \rightarrow {\bf 8} +{\bf 3} +\bar{\bf 3}$ for the
adjoint and ${\bf 7} \rightarrow {\bf 1} + {\bf 3} +\bar{\bf 3}$ for the
fundamental. The fundamental ${\bf 3}$ and $\bar{\bf 3}$ hypermultiplets cancel
the corresponding vector multiplets and leave a pure $SU(3)$ theory. Again we
have the problem that the curve is not Weyl invariant. The $SU(3)$ curve is
based on the fundamental ${\bf 3}$ of $SU(3)$. Under the $G_2$ Weyl group the
${\bf 3}$ is transformed into the $\bar{\bf 3}$. In fact, the $SU(3)$ curve is
simply reflected through the origin.

\subsection{$E_{6}$ with two fundamentals}

Finally we give some results for the group $E_{6}$. There is a regular
embedding of $SO(10)$ in $E_6$. In fact, the adjoint of $E_6$  breaks like $
{\bf 78} \rightarrow {\bf 45} + {\bf 16} + \bar{\bf  16} + \bf{1}$. If we add a
fundamental and an anti-fundamental hypermultiplet to the $E_6$ theory which
break like ${\bf 27} \rightarrow {\bf 16} + {\bf 10} +\bf{1}$  (and the
corresponding pattern for the anti-fundamental),  we obtain $SO(10)$ with two
fundamental hypermultiplets. The check of Dynkin indices also works out: $24 -
2 \times 6 = 16 -2 \times 2$. Unfortunately the proposed curve can only
describe a codimension 1 subspace of the $E_6$ moduli space.

We might note that the $F_4$ curve above also describes parts of the $E_6$
moduli space, since there is a special embedding of $F_4$ in $E_6$.  In that
case only a codimension 2 subspace is accessible.

\section{Some special examples}

\subsection{$SO(6) \rightarrow SO(5)$}

The adjoint representation of $SO(6)$ has dimension 15. The weights of the
adjoint correspond to the 3 Cartan elements and the 12 roots. The weight
diagram, with weights labelled by their Dynkin labels, see e.g. \cite{F}, is
given in fig 3.
\begin{figure}
\hbox to\hsize{\hss
\epsfysize=8cm
\epsffile{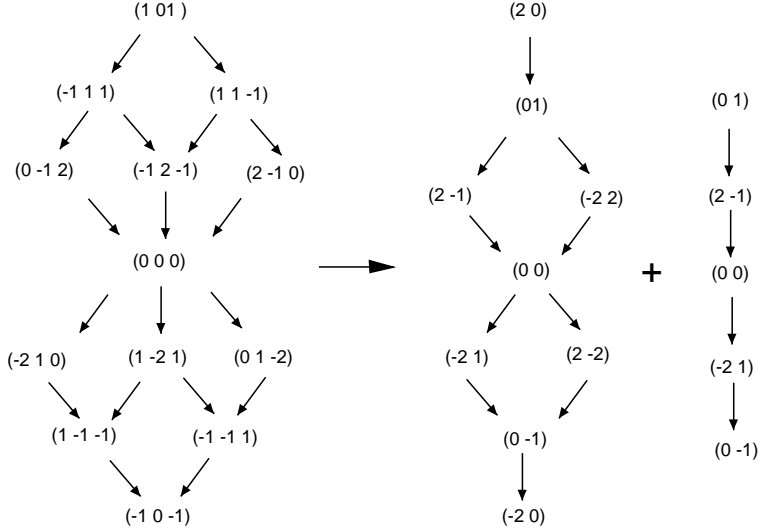}\hss}
\caption{The breaking $\bf{15} \rightarrow \bf{10} + \bf{5}$ of $SU(4)$ to
$SO(5)$.}
\end{figure}
The $(mass)^2$ of the gauge bosons are then proportional to $( \balpha \cdot
{\bf b})^2$. Let us now finetune the Higgs expectation value by setting $b_1 =
b_3$. We furthermore introduce  $\tilde{b} _1 = { b_1 + b_3 \over 2}$ and
$\tilde{b}_2 = b_2$.
A new, reduced weight diagram can now be constructed through $\balpha \cdot
{\bf b} \rightarrow {\bf w} \cdot \tilde{\bf b}$. It is drawn on the right of
fig. 3 where one can identify the $\bf 10$ (adjoint) and $\bf 5$ (fundamental)
of $SO(5)$. From the $SO(5)$ point of view we have vector-multiplets both in
the adjoint and the fundamental of $SO(5)$. Clearly this theory only makes
sense thanks to the embedding in $SO(6)$.

\begin{figure}
\hbox to\hsize{\hss
\epsfysize=6cm
\epsffile{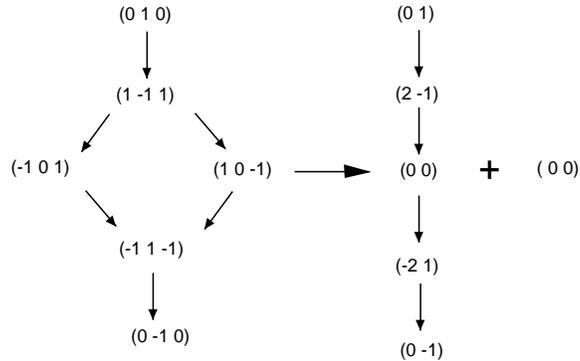}\hss}
\caption{The breaking $\bf{6} \rightarrow \bf{5} + \bf{1}$ of $SU(4)$ to
$SO(5)$.}
\end{figure}
Let us now add matter in the $\bf 6$ to the $SO(6)$ theory. We can also think
of this as an $SU(4)$ theory where we add matter in a tensor representation.
The $\bf 6$ breaks to a $\bf{5} + \bf{1}$ according to fig. 4. The
hypermultiplet will have charges and hence masses identical to the vector
multiplet in the fundamental representation. It follows then from
(\ref{verkan}) that the fundamental hypermultiplet will cancel the contribution
of the fundamental vector multiplet and leave a pure $SO(5)$ theory. We should
also check that the Dynkin indices come out correctly. For $SO(6)$ we have
$8-2=6$ which agrees with pure $SO(5)$.

\subsection{Pure $G_{2}$}

Let us now use the above ideas to work out the details for an exceptional
example, $G_2$. The adjoint, i.e. the $\bf 21$, of $SO(7)$ breaks to $\bf{14} +
\bf{7}$ of $G_2$. In the same way, the fundamental $\bf 7$ of $SO(7)$ goes to
the fundamental $\bf 7$ of $G_2$. This  means that the contribution of the
extra vector multiplets in the $G_2$ picture can be cancelled by adding a
hypermultiplet in the fundamental representation of $G_2$. We therefore
conclude that the pure $G_2$ theory can be obtained by restricting the Higgs
expectation values of an $SO(7)$ theory with a fundamental hypermultiplet. This
is also the same as a pure $SO(6)$ theory. The Dynkin indices work out as
$10-2=8$.

Let us check this in more detail! In fig. 5 we have drawn a choice of cycles
for $SO(7)$.
\begin{figure}
\hbox to\hsize{\hss
\epsfysize=8cm
\epsffile{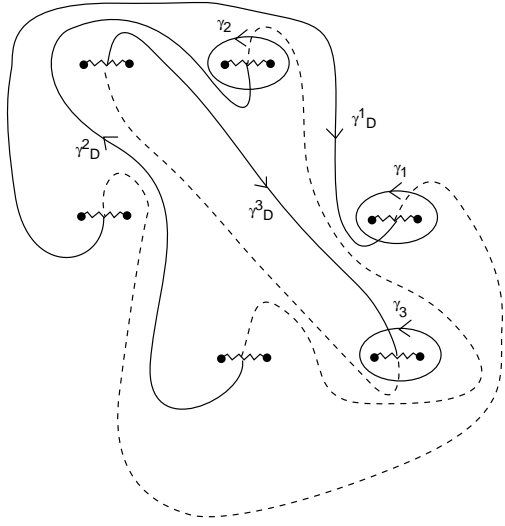}\hss}
\caption{Cycles for $SO(7)$ to be used for $G_2$.}
\end{figure}
The action of the three simple monodromies are shown in fig. 6. The
corresponding monodromy matrices are (in orthogonal basis)
\begin{eqnarray}
M_{1} = \pmatrix{
0 & 1 & 0 &  1 & -1  & 0\cr
1 & 0 & 0 & -1 & 1 & 0\cr
0 & 0 & 1 & 0 & 0 & 0\cr
0 & 0 & 0 &0 & 1 & 0\cr
0 & 0 & 0 &1 & 0 & 0\cr
0 & 0 & 0 & 0 & 0 & 1\cr
}
 \;\;,\;\;
M_{2} =\pmatrix{
1 & 0 & 0 &  0 & 0  & 0\cr
0 & 0 & 1 & 0 & 1 & -1\cr
0 & 1 & 0 & 0 & -1 & 1\cr
0 & 0 & 0 &1 & 0 & 0\cr
0 & 0 & 0 &0 & 0 & 1\cr
0 & 0 & 0 & 0 & 1 & 0\cr
}
, \nonumber \\
M_{3}=\pmatrix{
1 & 0 & 0 &  0 & 0  & 0\cr
0 & 1 & 0 & 0 & 0 & 0\cr
0 & 0 & -1 & 0 & 0 & 0\cr
0 & 0 & 0 &1 & 0 & 0\cr
0 & 0 & 0 &0 & 1 & 0\cr
0 & 0 & 0 & 0 & 0 & -1\cr
}   \hspace{3cm}
 \label{klassmonso7}
\end{eqnarray}

\begin{figure}
\hbox to\hsize{\hss
\epsfysize=8cm
\epsffile{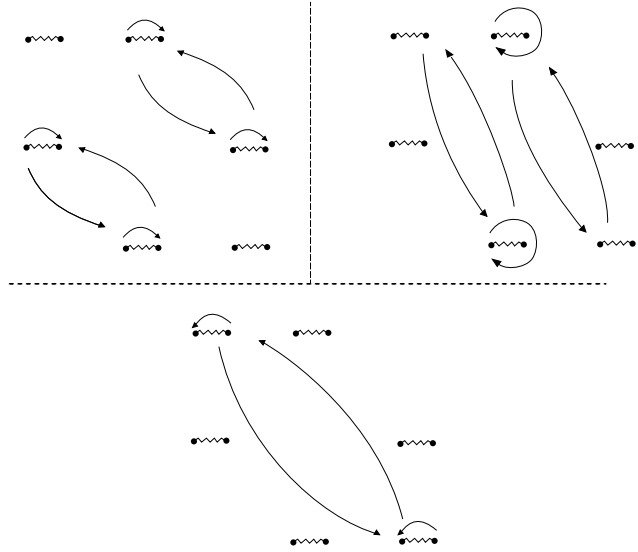}\hss}
\caption{Simple monodromies for $SO(7)$ with matter to be combined as $G_2$
monodromies.}
\end{figure}

To break to $G_2$ we must make a finetuning of the Higgs expectation values so
that we can write
\be
b_1 = \tilde{b}_2\;\;, b_2 = \tilde{b}_1 - \tilde{b}_2 , \;\; \mbox{and} \;\;
b_3 = 2 \tilde{b}_2 - \tilde{b}_1    , \label{bryt}
\ee
where $\tilde{b}_i$ are the Higgs expectation values in the $G_2$ theory. The
magnetic masses of the $G_2$ theory will be given by
$$
a_{1}^{D} = {1 \over 2} ({i \over 2 \pi} \oint _{\gamma _{2}^{D}} \lambda  -{i
\over 2 \pi} \oint _{\gamma _{3}^{D}} \lambda )
$$
\be
a_{2}^{D} = {1 \over 2} ({i \over 2 \pi}\oint _{\gamma _{1}^{D}} \lambda   -{i
\over 2 \pi}\oint _{\gamma _{2}^{D}}
\lambda ) +{i \over 2 \pi}\oint _{\gamma _{3}^{D}} \lambda .
\ee
The expressions for the magnetic cycles follow from $a_{i}^{D} = \frac{\partial
{\cal F}}{\partial a_{i}}$. We have chosen Dynkin basis for the $G_2$ group.
With these definitions we can write down the two simple monodromies of $G_2$ as
\begin{eqnarray}
M_{1} M_{3}= \pmatrix{
1 & 1 & -1 & 2 \cr
0 & -1 & 2 & -4 \cr
0 & 0 & 1 & 0 \cr
0 & 0 & 1 & -1 \cr
}
 , &
M_{2} =\pmatrix{
-1 & 0 & -4 & 6 \cr
3 & 1 & 6 & -9 \cr
0 & 0 & -1 & 3 \cr
0 & 0 & 0 & 1 \cr
} , \label{klassmon}
\end{eqnarray}
which is precisely what to be expected. We recall the general formula obtained
in \cite{var};
\be
	M_k \equiv
	 \pmatrix{
		W^t &   -\balpha_k \otimes \balpha_k \cr
					0 & W \cr
			}   .
	\label{simple mon}
\ee
We conclude that the pure $G_2$ theory is described by
\be
y^2 =((x^2-b_1 ^2)(x^2 - b_2 ^2 )(x^2 - b_3 ^2 ))^2 - x^4 \Lambda ^8 .
\ee
where $b_1$,  $b_2$  and $b_3$ are given by (\ref{bryt}).  One can note that
the curve is based on the fundamental weight diagram of $G_2$.

\section{An odd example}

Let us end with an equivalence that does not comfortably fit in the two classes
discussed above. This illustrates that maps between weght diagrams are the
primary objects in the equivalences, and subgroups just give a natural way to
generate these maps. We will consider $SU(4) \rightarrow SO(5)$, where we add
two hypermultiplets in the $\bf{4}$ to the $SU(4)$ theory. Under the above
breaking the $\bf{4}$ breaks like in fig. 7.
\begin{figure}
\hbox to\hsize{\hss
\epsfysize=6cm
\epsffile{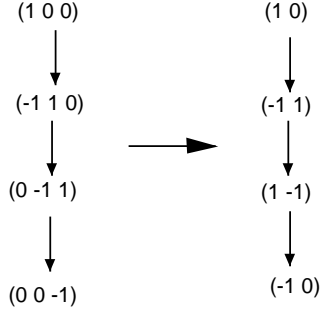}\hss}
\caption{The breaking $\bf{4} \rightarrow \bf{4}$ of $SU(4)$ to $SO(5)$ (or
$Sp(4)$).}
\end{figure}
Let us add these contributions to the broken adjoint of fig. 3 and write down
the prepotential. It is given by
$$
{\cal F} \sim {i \over 4 \pi} \bigg(
2 ( {\bf b} \cdot \balpha _{1} )^2 \log {({\bf b} \cdot \balpha _{1} )^2  \over
\Lambda ^2}
+ ( {\bf b} \cdot \balpha _{2} )^2 \log {({\bf b} \cdot \balpha _{2} )^2  \over
\Lambda ^2}
$$
$$
+ 2( {\bf b} \cdot (\balpha _{1} +\balpha _{2}) )^2 \log {({\bf b} \cdot
(\balpha _{1} + \balpha _{2}) )^2  \over
\Lambda ^2}
+ ( {\bf b} \cdot (2 \balpha _{1} +\balpha _{2}) )^2 \log {({\bf b} \cdot
(2\balpha _{1} + \balpha _{2}))^2  \over
\Lambda ^2}
$$
\be
-2( {\bf b} \cdot \balpha _{2}/2 )^2 \log {({\bf b} \cdot \balpha _{2} /2 )^2
\over
\Lambda ^2}
-2 ( {\bf b} \cdot (2 \balpha _{1} +\balpha _{2})/2 )^2 \log {({\bf b} \cdot
(2\balpha _{1} + \balpha _{2})/2)^2  \over
\Lambda ^2} \bigg)   .
\ee
The factor $2$ in the first and third term is due to the extra vector
multiplets we discussed in the previous subsection, the factor $2$ in the last
two terms is present since there are two hypermultiplets. Note also the factor
$1/2$ in the charge of the hypermultiplets. One can easily check that the
prepotential of $SO(5)$ is obtained after a renaming of the long and short
roots, i.e. $\balpha _{2} \rightarrow \sqrt{2} \balpha _{1}$ and
$\balpha _{1} \rightarrow  {1 \over \sqrt{2}} \balpha _{2}$.
This indeed results in the prepotential of pure $SO(5)$. The check of Dynkin
indices again works out: $8 - 2 \times 1 = 6$. The curves, according to
equations (5) and (7) are identical.

\section{Conclusions}

In this paper we have described some equivalences between different $N=2$
supersymmetric Yang-Mills theories. We have found all equivalences based on
maximal simple subgroups of simple Lie groups, and we have also observed in an
example that more general constructions are possible. We have used these
equivalences to construct  solutions of some theories with exceptional gauge
groups.  It is interesting to note  that  in some cases we need to relax  the
requirement of Weyl invariance of the complex curves. In effect, part of the
large Weyl symmetry in these theories is hidden. This allows hyperelliptic
curves to describe a larger set of theories. It is our hope that these ideas
can be of help also in a more general context.

\section*{Acknowledgements}

We wish to thank G. Ferretti and P. Stjernberg for discussions.

\end{document}